\documentclass[12pt,a4paper]{article}
\usepackage[english]{babel}
\usepackage{graphicx}

\newcommand{\alt}{\mathbin{\lower 3pt\hbox
   {$\rlap{\raise 5pt\hbox{$\char'074$}}\mathchar"7218$}}}
\newcommand{\agt}{\mathbin{\lower 3pt\hbox
   {$\rlap{\raise 5pt\hbox{$\char'076$}}\mathchar"7218$}}}

\begin{document}
\setcounter{footnote}{0}
\setcounter{equation}{0}
\setcounter{figure}{0}
\setcounter{table}{0}
\vspace*{5mm}

\begin{center}
{\large\bf Analytical asymptotics of
$\beta$-function in $\varphi^4$ theory\\
(end of the "zero charge" story)}

\vspace{4mm}
\vspace{4mm}
I. M. Suslov \\
P.L.Kapitza Institute for Physical Problems,
\\ 119337 Moscow, Russia \\
\vspace{1mm}
\end{center}

\begin{center}
\begin{minipage}{135mm}
{\bf Abstract } \\
Reconstruction of the $\beta$-function for $\varphi^4$ theory,
attempted  previously by summation of perturbation series, led
to the asymptotics $\beta(g)=\beta_\infty g^\alpha$ at
$g\to\infty$, where $\alpha\approx 1$ for space dimensions
$d=2,3,4$. The natural hypothesis arises, that asymptotic
behavior is $\beta(g) \sim g$ for all $d$.  Consideration of the
"toy" zero-dimensional model confirms the hypothesis and reveals
the origin of this result: it is related with a zero of a certain
functional integral.  Consideration can be generalized to the
arbitrary space dimensionality, confirming the linear asymptotics
of $\beta(g)$ for all $d$.  Asymptotical behavior for other
renormalization group functions (anomalous dimensions) is found
to be constant.  Relation to the "zero charge" problem is
discussed.

\end{minipage}
\end{center}


\vspace{6mm}
\begin{center}
{\bf 1. Introduction}
\end{center}

According to Landau, Abrikosov, Khalatnikov \cite{1},
relation of the bare charge $g_0$ with observable
charge $g$ for renormalizable field theories is
given by expression
$$
g=\frac{g_0}{1+\beta_2 g_0 \ln \Lambda/m}  \,,
\eqno(1)
$$
where $m$ is the mass of the particle, and $\Lambda$ is
the momentum cut-off.  For finite $g_0$ and
$\Lambda\to \infty$  the "zero charge"
situation ($g\to 0$) takes place. The proper interpretation of Eq.1
consists in its inverting, so that $g_0$  (related to the
length scale $\Lambda^{-1}$)
is chosen to give a correct value of $g$:
$$
g_0=\frac{g}{1-\beta_2 g \ln \Lambda/m} \,.
\eqno(2)
$$
The growth of $g_0$ with  $\Lambda$ invalidates Eqs.1,2
in the region $g_0\sim 1$
and existence of "the Landau pole" in Eq.2 has no physical sense.

The actual behavior of the charge  $g(L)$ as a function of the
length  scale $L$ is determined by the Gell-Mann -- Low equation
$$
-\frac{dg}{d \ln L} =\beta(g)=\beta_2 g^2+\beta_3 g^3+\ldots
\eqno(3)
$$
and depends on the appearance of the function  $\beta(g)$.
According to classification by Bogolyubov and Shirkov \cite{2},
the growth of  $g(L)$ is saturated, if $\beta(g)$ has a zero for
finite  $g$, and continues to infinity, if $\beta(g)$ is
non-alternating and behaves as $\beta(g)\sim g^\alpha$ with
$\alpha\le 1$ for large $g$; if, however, $\beta(g)\sim g^\alpha$
with $\alpha>1$, then $g(L) $ is divergent  at finite $L=L_0$
(the real Landau pole arises) and the theory is internally
inconsistent due to indeterminacy of $g(L)$ for $L<L_0$. The
latter case corresponds to the "zero charge" situation in full
theory (beyond its perturbation framework).

One can see that solution of the "zero charge" problem needs
calculation of the Gell-Mann -- Low function $\beta(g)$ at
arbitrary $g$, and in particular its asymptotic behavior for
$g\to\infty$. Such attempt was made recently by the present author
for $\varphi^4$ theory \cite{4}, QED \cite{5} and QCD \cite{6}
(see a review article \cite{7}). It is based on the fact that the
first four coefficients $\beta_N$ in Eq.3 are known from
diagrammatic calculations, while their large order behavior can be
established by the Lipatov method \cite{10,7}. Smooth
interpolation of the coefficient function and the proper summation
of the perturbation series give non-alternating $\beta(g)$ with
$\alpha\approx 1$ in four-dimensional $\varphi^4$ theory
\cite{4}\,\footnote{\,Possibility of correct summation of perturbation
series is frequently questioned in relation with possible
existence of renormalon singularities in the Borel plane \cite{200}.
Such  singularities can be easily obtained by summing some special
sequences of  diagrams, but their existence was never
proved, if all diagrams are taken into account \cite{201}.
The present result for the
asymptotics of $\beta(g)$ satisfies the general criterion for
absence of renormalon singularities \cite{51} and confirms
the proof of their absence  suggested in
\cite{202}.}.
Recent results for 2D and 3D  $\varphi^4$ theory  \cite{14,15}
also correspond to  $\alpha\approx 1$. The natural hypothesis
arises, that $\beta(g)$ has the linear asymptotics for arbitrary
space dimension  $d$. Simplicity of the result indicates that it
can be obtained analytically.

Below we show that it is so indeed. Analysis of
zero-dimensional theory confirms the asymptotics $\beta(g)\sim g$
and reveals its origin. It
is related with unexpected
circumstance that the
 strong coupling limit for the renormalized charge  $g$
 is determined not by large values of the bare
charge  $g_0$,  but its complex values.\,\footnote{\,Discussion
of unitarity of theory is given in
Sec.5.}
More than that, it is
sufficient to consider the region $|g_0|\ll 1$, where the
functional integrals can be evaluated in the saddle-point
approximation. If a proper direction in the complex $g_0$ plane is
chosen, the saddle-point contribution of the trivial vacuum is
comparable with the saddle-point contribution of the main
instanton, and a functional integral can turn to zero. The limit
$g\to\infty$ is related with the zero of a certain functional
integral and appears to be completely controllable.  As a result,
it is possible to obtain asymptotic behavior of the
$\beta$-function and anomalous dimensions: the former indeed
appears to be linear, while the latter achieve certain constant
limits. Analogous results can be obtained for QED \cite{3111}.

The asymptotics $\beta(g)\sim g$ in combination with
non-alternating
behavior of $\beta(g)$ corresponds to the second possibility in
the Bogolyubov--Shirkov classification: $g(L)$ is finite for large
$L$ but grows to infinity at $L\to 0$. It looks in conflict with
the expected triviality of  $\varphi^4$ theory (see e.g. \cite{41}
and the references therein). In fact, two definitions of
triviality were mixed in the literature. The first one, introduced
by Wilson \cite{21}, is equivalent to positiveness of $\beta(g)$
for $g\ne 0$; it is confirmed by all available information and can
be considered as firmly established. The second definition,
introduced by mathematical
community  \cite{30}, corresponds to the true triviality and is
equivalent to internal inconsistency in the Bogolyubov--Shirkov
sense: it needs not only positiveness of $\beta(g)$ but also the
corresponding asymptotical behavior. Evidence of true triviality
is not extensive
and allows different interpretation \cite{4}. The present
analysis gives new insight to this problem: to obtain nontrivial
theory one need to use the complex values of the bare charge $g_0$,
which were never exploited in mathematical proofs and numerical
simulations.  This matters will be discussed in a
separate paper \cite{31}.

\vspace{6mm}
\begin{center}
{\bf 2. Definition of the renormalization group (RG) functions}
\end{center}

Consider the $O(n)$ symmetric  $\varphi^4$ theory with an
action
$$
S\{\varphi\} =\int \,d^dx \left\{
{\textstyle\frac{1}{2}} \sum_\alpha (\nabla \varphi_\alpha)^2
+ {\textstyle\frac{1}{2}} m_0^2 \sum_\alpha \varphi_\alpha^{\,2} +
{\textstyle\frac{1}{8}} u
\left(\sum_\alpha\varphi_\alpha^{\,2}\right)^2 \right\}\,,
$$
$$
u=g_0\Lambda^{\epsilon}\,, \qquad \epsilon=4-d
\eqno(4)
$$
in $d$--dimensional space. Following the usual RG
formalism \cite{16}, consider the "amputated" vertex
$\Gamma^{(L,N)}$
with $N$ external lines of field $\varphi$ and  $L$
insertions of $\varphi^2$ (where two fields $\varphi$ are
taken in the coinciding spatial points).
Its multiplicative
renormalizability  means
$$
\Gamma^{(L,N)}(p_i; g_0, m_0, \Lambda) = Z^{-N/2}
\left(\frac{Z_2}{Z} \right)^{-L}
\Gamma_R^{(L,N)}(p_i; g,m)\,,
\eqno(5)
$$
i.e. divergency at $\Lambda\to\infty$ disappears after
extracting the proper $Z$-factors and transferring to the
renormalized charge and mass. We accept the renormalization
conditions at zero momentum\,\footnote{\,Dependence on the
renormalization scheme is discussed in Sec.5.}
$$
\left.\Gamma_R^{(0,2)}(p;
g,m)\right|_{p\to 0} =m^2 + p^2 + O(p^4)\,, \qquad
$$
$$
\left.\Gamma_R^{(0,4)}(p_i; g,m)\right|_{p_i=0} =g m^\epsilon\,,
\eqno(6)
$$
$$
\left.\Gamma_R^{(1,2)}(p_i; g,m)\right|_{p_i=0}
=1\,,
$$
which are typical for applications in the phase transitions
theory \cite{16a}. Substitution of (6) into (5) gives
expressions for $g$, $m$, $Z$, $Z_2$  in terms of the bare
quantities
$$
Z(g_0, m_0, \Lambda) = \left( \frac{\partial}{\partial p^2}
\left. \Gamma^{(0,2)}(p; g_0, m_0, \Lambda) \right|_{p=0}
\right)^{-1} \,,
$$
$$
Z_2(g_0, m_0, \Lambda) = \left(
\left. \Gamma^{(1,2)}(p_i; g_0, m_0, \Lambda) \right|_{p_i=0}
\right)^{-1}    \,,
\eqno(7)
$$
$$
m^2=Z(g_0, m_0, \Lambda)
\left. \Gamma^{(0,2)}(p; g_0, m_0, \Lambda)
\right|_{p=0} \,,
$$
$$
g = m^{-\epsilon} Z^2(g_0, m_0, \Lambda)
\left. \Gamma^{(0,4)}(p_i; g_0, m_0, \Lambda)
\right|_{p_i=0}  \,.
$$
Applying differential operator $d/d\ln m$ to (5) for fixed $g_0$
and $\Lambda$  gives the Callan-Symanzik equation,
valid asymptotically for large  $p_i/m$ \cite{16}
$$
\left[ \frac{\partial}{\partial\ln m} +
\beta(g) \,\frac{\partial}{\partial g}
+\left(L-N/2 \right) \eta(g) -L\eta_2(g)
\right] \Gamma^{(L,N)}_R(p_i; g,m) \approx 0\,,
\eqno(8)
$$
where the RG functions $\beta(g)$, $\eta(g)$ and
$\eta_2(g)$ are determined as
$$
\beta(g)=\left. \frac{d g}{d \ln m} \right|_{g_0,\Lambda=const}
\,,\qquad
\eta(g)=\left. \frac{d \ln Z}{d \ln m} \right|_{g_0,\Lambda=const}
\,,\qquad
\eta_2(g)=\left. \frac{d \ln Z_2}{d \ln m}
\right|_{g_0,\Lambda=const}
\eqno(9)
$$
and according to general theorems depend only on  $g$ \cite{16}.

\vspace{6mm}
\begin{center}
{\bf 3. "Naive" zero-dimensional limit.}
\end{center}

The functional integrals of  $\varphi^4$ theory are determined
as
$$
Z^{(M)}_{\alpha_1\ldots \alpha_M}(x_1,\ldots, x_M)=
\int D\varphi\,
\varphi_{\alpha_1} (x_1) \varphi_{\alpha_2} (x_2) \ldots
\varphi_{\alpha_M} (x_M) \exp\left(-S\{\varphi \} \right) \,.
\eqno(10)
$$
To take a zero-dimensional limit, consider the system restricted
spatially in all directions at sufficiently small scale, and
neglecting spatial dependence of  $\varphi(x)$ omit the terms
with gradients in Eq.4; interpreting the functional integral as
a multi-dimensional integral on a lattice, we can take the system
sufficiently small, so it contains only one lattice site:
$$
Z^{(M)}_{\alpha_1\ldots \alpha_M} =
\int d^n\varphi\,
\varphi_{\alpha_1} \ldots
\varphi_{\alpha_M}
\exp\left(-{\textstyle\frac{1}{2}} m_0^2 \varphi^2-
{\textstyle\frac{1}{8}} u \varphi^{4}   \right) \,.
\eqno(11)
$$
The diagrammatic expansions generated by such "functional"
integrals have the usual form, but all propagators should be
taken at zero momenta and no momentum integrations are necessary.

Such understanding of zero-dimensional theory is
conventional in the literature. However, it does not quite
correspond to the true zero-dimensional limit of $\varphi^4$
theory.  Considering expressions for the simplest diagrams in
$d$-dimensional case and taking limit $d\to 0$, it it easy to be
convinced that their trivialization (of the described type)
occurs only for zero external momenta; if the latter are
different from zero, no evident simplifications occur. This
point is essential for definition of the  $Z$-factor, which is
introduced according to a scheme (see the first relation in (6))
$$
G_2(p)=\frac{1}{p^2+m_0^2+\Sigma(p,m_0)} =
\frac{1}{p^2+m_0^2+a_0(m_0)+a_2(m_0)p^2+a_4(m_0)p^4+\ldots} =
$$
$$
=\frac{Z}{p^2+m^2+O(p^4)} \,,
\eqno(12)
$$
and is determined by the momentum dependence of self-energy.
In the described "naive" zero-dimensional theory, non-zero
momenta are absent and we can accept  $Z=1$. Such procedure is
internally consistent but does not correspond to the true
zero-dimensional limit of $\varphi^4$ theory. The latter fact
is not  essential for us, since this model is used only for
illustration and the proper consideration of the general
$d$-dimensional case will be given in the next section.

Substituting  $\varphi_\alpha=\varphi u_\alpha$ in (11)
 and integrating over directions of the unit vector
${\bf u}$, we obtain for even $M$ \cite{17}
$$
Z^{(M)}_{\alpha_1\ldots \alpha_M}=
\frac{2\pi^{n/2}}{2^{M/2} \Gamma(M/2+n/2)}
I_{\alpha_1 \ldots  \alpha_M} \, K_M(m_0,u)\,,
\eqno(13)
$$
where $I_{\alpha_1 \ldots  \alpha_M}$ is the sum of terms
like  $\delta_{\alpha_1 \alpha_2} \delta_{\alpha_3
\alpha_4}\ldots$ with all possible pairings, and
$$
K_M(m_0,u) = \int_0^\infty \varphi^{M+n-1} d\varphi\,
\exp\left(-{\textstyle\frac{1}{2}} m_0^2 \varphi^2-
{\textstyle\frac{1}{8}} u \varphi^{4}   \right) \,.
\eqno(14)
$$
Defining the $M$--point Green functions $G^{(M)}$ as
$Z^{(M)}/Z^{(0)}$ and extracting dependence on indices
$$
G^{(2)}_{\alpha \beta} =G_2 \delta_{\alpha
\beta}\,, \qquad G^{(4)}_{\alpha \beta \gamma \delta} = G_4
I_{\alpha \beta \gamma \delta} \,, \qquad \Gamma^{(0,4)}_{\alpha
\beta \gamma \delta} = \Gamma_4 I_{\alpha \beta \gamma \delta}
\,,
\eqno(15)
$$
we have
$$
\Gamma_2=1/G_2\,,\qquad
G_4=G_2^2-G_2^4 \Gamma_4 \,, \eqno(16)
$$
where
$$
G_2=\frac{1}{n} \frac{K_2(m_0,u)}{K_0(m_0,u)}\,, \qquad
G_4=\frac{1}{n(n+2)} \frac{K_4(m_0,u)}{K_0(m_0,u)}\,
\eqno(17)
$$
and the vertex  $\Gamma^{(0,4)}_{\alpha \beta \gamma \delta}$
is defined by the usual relation
$$
G^{(4)}_{\alpha \beta \gamma \delta} =
G^{(2)}_{\alpha \beta} G^{(2)}_{\gamma \delta} +
G^{(2)}_{\alpha \gamma} G^{(2)}_{\beta \delta}
+ G^{(2)}_{\alpha \delta} G^{(2)}_{\beta \gamma} -
G^{(2)}_{\alpha \alpha'} G^{(2)}_{\beta \beta'}
G^{(2)}_{\gamma \gamma'} G^{(2)}_{\delta \delta'}
\Gamma^{(0,4)}_{\alpha' \beta' \gamma' \delta'}  \,.
\eqno(18)
$$
Using the renormalization conditions (7), we obtain
$$
m^2=\Gamma_2=\frac{nK_0}{K_2} \,,
\eqno(19)
$$
$$
g=\frac{\Gamma_4}{m^4} =  1-m^4 G_4=
1- \frac{n}{n+2} \frac{K_4 K_0}{K_2^2} \,.
\eqno(20)
$$
Differentiating (19) over $m_0^2$ and taking into account
that differentiation of $K_M$ transfers it to
$K_{M+2}$ (see  (14)), we have
$$
\frac{d m^2}{d m_0^2} = \frac{n}{2}
\left\{ -1+\frac{K_4 K_0}{K_2^2} \right\} \,.
\eqno(21)
$$
Since all differentiations in (9) occur at
$g_0,\,\,\Lambda = const$, the latter parameters are considered
to be fixed throughout all calculations: then $m^2$ is a
function of only $m_0^2$ and Eq.21 defines also the derivative
$d m_0^2/d m^2$. According to definition of the
$\beta$-function (9) we have
$$
\beta(g)=2 \frac{d g}{d\ln m^2}=
-\frac{2 m^4}{n(n+2)} \left[\, 2\, \frac{K_4}{K_0} +
\left( \frac{K_4}{K_0} \right)'_{m_0^2} \,m^2\,
\frac{d m_0^2}{d m^2} \right]
\eqno(22)
$$
and substitution of (21) gives the following expression
$$
\beta(g)=-\frac{2n}{n+2} \frac{K_4 K_0}{K_2^2}
\left[ 2+\frac{ \frac{K_6 K_0}{K_4 K_2}-1 }
{ 1-\frac{K_4 K_0}{K_2^2}} \right] \,.
\eqno(23)
$$
The change of variables $\varphi\to \varphi (8/u)^{1/4} $
in the integrals (14) reduces them to the form
$$
K_M(t) = \int_0^\infty \varphi^{M+n-1} d\varphi\,
\exp\left(-t \varphi^2- \varphi^{4}   \right) \,,
\qquad t=\left(\frac{2}{u} \right)^{1/2} \,m_0^2 \,.
\eqno(24)
$$
The arising factors drop out of the combinations
$K_4 K_0/K_2^2$ and $K_6 K_0/K_4 K_2$ entering
equations (20),(23), and the latter
have the same form in terms of $K_M(t)$, as
they had in terms of $K_M(m_0,u)$.
The right hand
sides of (20),(23) are the functions of the single
variable $t$ and the dependence $\beta(g)$ is determined
by these expressions in the parametric form.

The vertex
$\Gamma^{(1,2)}_{\alpha \beta}=\Gamma_{12}\delta_{\alpha\beta}$
is determined by the Ward identity,
$$
\Gamma_{12}=\frac{d m^2}{d m_0^2}=1- \frac{n+2}{2}\,g \,,
\eqno(25)
$$
and the function $\eta_2(g)$ is given by expression
$$
\eta_2(g)=- \frac{d\ln \Gamma_{12}}{d\ln m} =
\frac{\beta(g)}{2/(n+2)-g} \,,
\eqno(26)
$$
while $\eta(g)$ is identically zero in the accepted
approximation.

\vspace{2mm}

Using the asymptotic expressions for $K_M(t)$ \,,
$$
K_M(t)= \left\{ \begin{array}{cc}
\frac{1}{\sqrt{2}} t^{-(M+n)/2} \Gamma\left(\frac{M+n}{2}\right)
\left[ 1-\frac{(M+n)(M+n+2)}{4t^2}+\ldots \right]\,, &
                                                t\to\infty\\
{ }\\
\frac{1}{4} \left[ \Gamma\left(\frac{M+n}{4}\right)
- t \Gamma\left(\frac{M+n+2}{4}\right)+\ldots \right]\,, &t\to 0
\\ {   }\\
\frac{\sqrt{\pi}}{2} {\rm e}^{t^2/4}
\left(\frac{|t|}{2}\right)^{(M+n-2)/2} \left[
1+\frac{(M+n-2)(M+n-4)}{4t^2}+\ldots \right]  \,, &
t \to -\infty \,,
\end{array} \right.\,,
\eqno(27)
$$
it is easy to obtain that  $g$ and $\beta(g)$ depend on
$t$ as shown in Fig.\,1,a,
\begin{figure}
\centerline{\includegraphics[width=5.1 in]{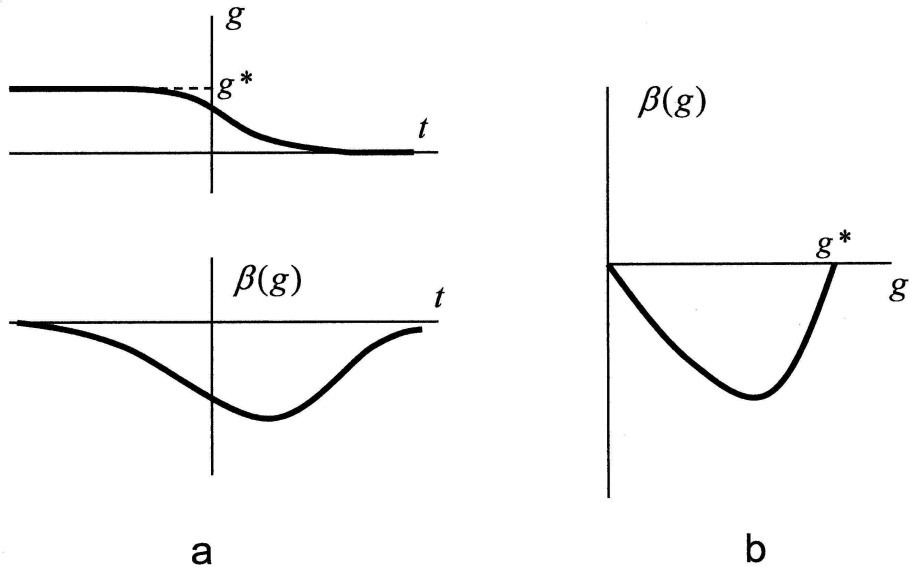}} \caption{
(a) Dependence of $g$ and $\beta(g)$ on the parameter $t$. (b)
Resulting appearance of $\beta(g)$. } \label{fig1}
\end{figure}
i.e. variation of parameter $t$ along the real axis determines
$\beta(g)$ in the interval from $g=0$ till the fixed point
(Fig.\,1,b)\,\footnote{\,Existence of the fixed point $g^*$ does
not mean the existence of the phase transition, which is absent
for  $d<2$.  The scaling behavior of correlation functions follows
from the Callan--Symanzik equation only in the region of small
$m$, which is inaccessible for physical values of $m_0$ and $g_0$.
Eq.(28) is in agreement with the result  $\tilde g^*=(n+8)/(n+2)$,
obtained in \cite{20a}, where normalization of charge $\tilde g$
differs from our, $\tilde g=(n+8)g/2$.  As discussed above, this
result does not correspond to the true zero-dimensional limit of
$\varphi^4$ theory and its use in the interpolation scheme for
improving dependence of $g^*$ on the space dimension $d$
\cite{20a} is not reasonable.}
$$
g^* = \frac{2}{n+2} \,.
\eqno(28)
$$
To advance into the large $g$ region, one should investigate
the parametric representation  (20), (23) for complex values of
$t$.  If $t=|t| {\rm e}^{i\chi}$ and $|t|\gg 1$, then (in
dependence on  $\chi$)  the integrals  $K_M(t)$ are determined by
either trivial saddle point   $\varphi=0$, or nontrivial
saddle-point   $\varphi^2=-t/2$. The saddle-point
contributions to $K_M(t)$ depend on $t$, but this dependence
drops out of the combinations $K_4 K_0/K_2^2$ and $K_6 K_0/K_4
K_2$, entering (20,23). Thus, in the rough approximation, the
complex  $t$ plane is divided into two parts where  $g$ and
$\beta(g)$ takes constant values $g=0$, $\beta(g)=0$
and $g=g^*$, $\beta(g)=0$. The smooth transition between these
values is related with deviations from the saddle-point
approximation, which arise for  $|t|\alt 1$; however,
corresponding variations of $g$ are expected to be finite, as in
the case of the real  $t$ (Fig.\,1,a). Now it is easy to understand
that large values of $g$ can be achieved only in those directions
of the complex $t$ plane where contributions from two saddle
points are comparable  in value. Then for  $K_M(t)$ we have
representation
$$
K_M(t)= A {\rm e}^{i\psi} + A_1 {\rm e}^{i\psi_1} =
A {\rm e}^{i\psi} \left( 1 + a {\rm e}^{i\Delta} \right)
\eqno(29)
$$
and the integral can be turned to zero by the corresponding
choice of $a$ and $\Delta$. Indeed, two available degrees of
freedom (${\rm Re}\,t$ and ${\rm Im}\,t$) are in principle
sufficient to adjust $a$ and $\Delta$. With variation of $t$,
parameter $a$ surely passes through the unit value, since the
complex $t$ plane contains regions where dominates either the
first, or the second term of (29). As for the change of $\Delta$,
it occurs in infinite limits (see below), and the integral
$K_M(t)$ has an infinite number of zeroes lying close to the lines
$\chi=\pm 3\pi/4$ and accumulating at infinity. Therefore, the
saddle-point approximation used in above considerations can be
justified for zeroes lying in the large $|t|$ region.

It is easy to see that the limit $g\to \infty$ can be
achieved, if $K_2$ goes to zero; then  (20,23) are
simplified,
$$
g\approx - \frac{n}{n+2} \frac{K_4 K_0}{K_2^2} \,,\qquad
\beta(g)\approx - \frac{4n}{n+2} \frac{K_4 K_0}{K_2^2}\,,
\eqno(30)
$$
and the parametric representation can be resolved in the form
$$
\beta(g)= 4 g\,, \qquad g\to\infty \,,
\eqno(31)
$$
while from (26) we have
$$
\eta_2(g) = -4\,, \qquad g\to\infty   \,.
\eqno(32)
$$
In accordance with expectations, the asymptotics of $\beta(g)$ appears
to be linear.

\vspace{2mm}

In derivation of  (31), (32) we did not use the explicit form
of the integrals  $K_M(t)$: it was essential only that (a) the
integral $K_2(t)$ has any zeroes, and (b) zeroes of different
integrals  $K_M(t)$ do not coincide. Let us show that it is indeed
so. The values of action for the saddle points  $\varphi=0$  and
$\varphi^2=-t/2$ are equal to  0 and $t^2/4$ correspondingly, and
contributions of these points are comparable for  $Re\, t^2 =0$ or
$\chi=\pm \pi/4, \pm 3\pi/4$. However, values $\chi=\pm \pi/4$ are
not suitable: the integral $K_M(t)$ exhibits the Stokes
phenomenon, which is related with the change of topology for lines
of the steepest descent (see, e.g. \cite{20}). This change of
topology  occurs at $\chi=\pm\pi/2$: for $0<|\chi|<\pi/2$ the line
of the steepest descent passes only the trivial saddle point
(Fig.\,2,a), while for $\pi/2<|\chi|<\pi$ both
\begin{figure}
\centerline{\includegraphics[width=5.1 in]{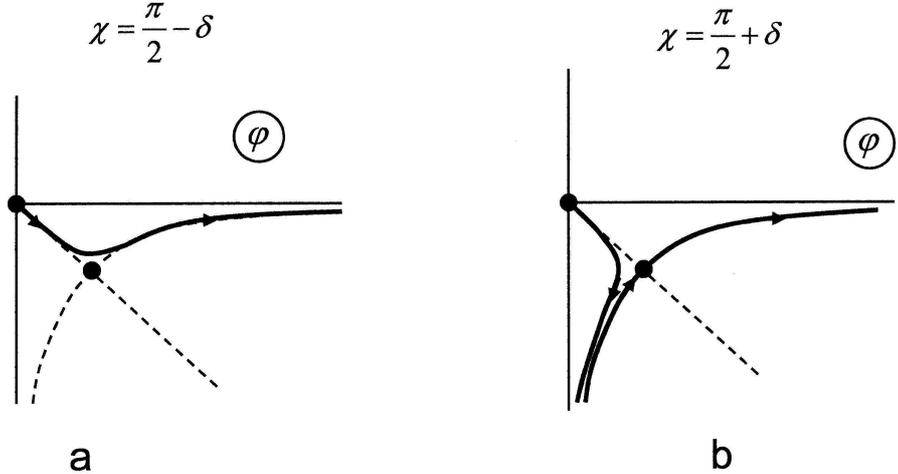}}
\caption{The lines of the steepest descent for the
integral $K_M(t)$: (a) for $0<\chi<\pi/2$, and
(b) for $\pi/2<\chi<\pi$. } \label{fig2}
\end{figure}
saddle points are passed (Fig.\,2,b). The compensation
of two contributions (29) is possible for $\chi=\pm 3\pi/4$,
but  does not occur for $\chi=\pm \pi/4$.  Setting
$t=\rho {\rm e}^{i\chi}$, $\rho\gg 1$, $\chi=3\pi/4+\delta$,
$\delta \ll 1$, we have for contributions of two saddle
points in the integral $K_0(t)$
$$
K_0(t)=\rho^{-n/2} {\rm e}^{-i\frac{3\pi}{8} n}
\left[ \frac{1}{2}
\Gamma\left(\frac{n}{2}\right) +\frac{\sqrt{\pi}}{2^{n/2}} {\rm
e}^{-i\frac{\pi}{4}+i\frac{\pi}{4} n -i\frac{\,1}{4}\rho^2}
\,\rho^{n-1} {\rm e}^{\frac{1}{2} \rho^2 \delta}
 \right]
\eqno(33)
$$
Choosing  $\delta(\rho)$ from the condition
$$
\rho^{n-1} {\rm e}^{\frac{1}{2} \rho^2 \delta}=
\frac{2^{n/2-1}}{\sqrt{\pi}} \Gamma\left(\frac{n}{2}\right)\,,
\qquad {\rm i.e.} \qquad \delta\sim \ln{\rho}/\rho^2
\eqno(34)
$$
we obtain
$$
K_0(t)= \frac{1}{2} \Gamma\left(\frac{n}{2}\right)
\rho^{-n/2} {\rm e}^{-i\frac{3\pi}{8} n}
\left[ 1+ {\rm e}^{\frac{i}{4}(\pi+\pi n-\rho^2)} \, \right]
\eqno(35)
$$
and zeroes of $K_0(t)$ lie in the points
$$
\rho_s^2=\pi(n+5) +8\pi s\,, \qquad s {\rm - integer} \,.
\eqno(36)
$$
The results for $K_M(t)$ can be obtained by the change
$n\to n+M$,  and it is clear from (34), (36) that
different integrals
$K_M(t)$ turn to zero in different points.

\vspace{4mm}

\begin{center}

{\bf 4. General $d$-dimensional case} \\
\end{center}
\vspace{3mm}

According to (24), the complex $t$  values with $|t|\to\infty$
correspond to complex  $g_0$ with $|g_0|\to 0$, and we come to
miraculous conclusion: large values of the renormalized charge $g$
corresponds not to large values of the bare charge $g_0$
(as naturally to think\,\footnote{\,It is commonly accepted
that some universal function $g=f(L)$ can be introduced,
describing the dependence of the charge on the length scale.
Then the observable charge corresponds to  $g_{obs}=f(m^{-1})$,
the bare charge corresponds to  $g_0=f(\Lambda^{-1})$,
and the renormalized charge defined at the scale  $L$, is simply
$g=f(L)$, i.e. all charges entering the theory are in fact
one and the same charge but related with different scales.
However, it is well-known that this picture is approximate
due to ambiguity of the renormalization scheme. Definitions
of the bare and renormalized charge are technically different
and introduced in the cut-off and subtraction schemes,
correspondingly \cite{20b}. Associated functions  $g_0=f_1(L)$
and $g=f_2(L)$ coincide on the one-loop and two-loop level,
but differ in higher orders. Hence, our intuition is relevant
only in the weak coupling region.  }), but to its complex values;
more than that, it is sufficient to consider the region
$|g_0|\ll 1$, where the saddle-point approximation is applicable.

As a result,  the zeroes of the
functional integrals can be obtained by compensation of the
saddle-point contributions of trivial vacuum and of the instanton
configuration with the
 minimal action\,\footnote{\,Contributions
of higher instantons contain additional
smallness for $|g_0|\ll 1$.}. The latter
contribution is well-studied in relation with calculation of
the Lipatov asymptotics and for $d<4$ is given by expression
(see, e.g.  \cite{24}, Eq.93)
$$
\left[Z^{(M)}_{\alpha_1\ldots \alpha_M}(p_1,\ldots,
p_M)\right]^{inst}=
i c_M (-g_0)^{-(M+r)/2} {\rm e}^{-S_0/g_0}
\langle \phi_c \rangle_{p_1}\ldots \langle \phi_c \rangle_{p_M}
 \,I_{\alpha_1 \ldots \alpha_M}
\eqno(37)
$$
and by somewhat more complicated expression for $d=4$.
Here  $\langle \phi_c \rangle_{p}$ is the Fourier transform
of the dimensionless instanton configuration $\phi_c(x)$,
$S_0$ is the corresponding action, $r$ is the number of zero
modes, and $c_M$ is a certain constant. Then for
$M=0,2,\ldots$ we have
$$
Z_0=1+i c_0(-g_0)^{-r/2} {\rm e}^{-S_0/g_0} \,,
$$
$$
Z^{(2)}_{\alpha\beta}(p,p')=\frac{\delta_{\alpha\beta}}{p^2+m_0^2}+
i c_2(-g_0)^{-(r+2)/2} {\rm e}^{-S_0/g_0} \langle \phi_c
\rangle_p^2  \, \delta_{\alpha\beta} \,,
\eqno(38)
$$
etc., where all contributions are normalized by a value $Z^{(0)}$
at $g=0$. Setting  $t^2=-S_0/g_0$, we come to expression of type
(33), which can be analyzed analogously. It is easy to be
convinced that different integrals  $K_M$ and their derivatives
over $m_0^2$ have zeroes in different points.

\vspace{3mm}

Now we need representation of RG functions
in terms of functional integrals. The Fourier transform of
(10) is
$$
Z^{(M)}_{\alpha_1\ldots \alpha_M}(p_1,\ldots, p_M)
{\cal N}\delta_{p_1+\ldots+p_M}=
\sum\limits_{x_1, \ldots, x_M}
Z^{(M)}_{\alpha_1\ldots \alpha_M}(x_1,\ldots, x_M)
{\rm e}^{i p_1 x_1+\ldots +i p_M x_M}
\eqno(39)
$$
where ${\cal N}$ is the number of sites on the lattice, which
is implied in definition of the functional integral.
For the choice of external momenta corresponding to the symmetric
point,  $p_i\cdot p_j=p^2(4\delta_{ij}-1)/3$, it is possible
to extract factors $I_{\alpha_1 \ldots \alpha_M}$ from $Z^{(M)}$,
in analogy with (13)
 $$
Z^{(0)}=K_0\,,\qquad
Z^{(2)}_{\alpha \beta}(p, -p)=K_2(p)\delta_{\alpha \beta}
\,,\qquad
Z^{(4)}_{\alpha \beta \gamma \delta}\{p_i\}=
K_4\{p_i\}
I_{\alpha \beta \gamma \delta}
\eqno(40)
$$
Introducing vertex  $\Gamma^{(0,4)}$ by relation
$$
G^{(4)}_{\alpha \beta \gamma \delta}(p_1,\ldots,p_4) =
G^{(2)}_{\alpha \beta}(p_1)
   G^{(2)}_{\gamma \delta}(p_3)\,{\cal N}\delta_{p_1+p_2} +
G^{(2)}_{\alpha \gamma}(p_1)
       G^{(2)}_{\beta \delta}(p_2)\,{\cal N}\delta_{p_1+p_3} +
$$
$$
+ G^{(2)}_{\alpha \delta}(p_1)
G^{(2)}_{\beta \gamma}(p_3)\,{\cal N}\delta_{p_1+p_4} -
G^{(2)}_{\alpha \alpha'}(p_1) G^{(2)}_{\beta \beta'}(p_2)
G^{(2)}_{\gamma \gamma'}(p_3) G^{(2)}_{\delta \delta'}(p_4)
\Gamma^{(0,4)}_{\alpha' \beta'\gamma' \delta'}(p_1,\ldots,p_4)
\,  \eqno(41)
$$
and extracting $I_{\alpha_1 \ldots \alpha_M}$, we have
$$
G^{(2)}_{\alpha \beta}(p, -p)=G_2(p)\delta_{\alpha \beta}
 \,,\qquad
G^{(4)}_{\alpha \beta \gamma \delta}\{p_i\}=
G_4\{p_i\}
I_{\alpha \beta \gamma \delta} \,,\qquad
\Gamma^{(0,4)}_{\alpha \beta \gamma \delta}\{p_i\}=
\Gamma_4\{p_i\}
I_{\alpha \beta \gamma \delta}
\eqno(42)
$$
For strictly zero momenta $p_i$, the relation of $G_4$ to
$\Gamma_4$ contains factors  ${\cal N}$, proportional to
volume of the system. It is more convenient to set $p_i\sim \mu$,
excluding special equalities like $p_1+p_2=0$, and choose
$\mu$ so that ${\cal L}^{-1}\alt \mu\ll m$, where lower bound
goes to zero in the limit of the infinite system size
${\cal L}$. Then
$$
G_4=\frac{K_4}{K_0}\,,\qquad
\Gamma_4=-\frac{G_4}{G_2^4} =-\frac{K_4 K_0^3}{K_2^4} \,,
\eqno(43)
$$
where the integrals are taken at zero momenta, and
$$
G_2=\frac{K_2(p)}{K_0}\,,\qquad
\Gamma_2(p)=\frac{1}{G_2(p)} =\frac{K_0}{K_2(p)}
\approx \frac{K_0}{K_2} +\frac{K_0 \tilde K_2}{K_2^2}\,p^2
\eqno(44)
$$
where we have written for small  $p$\,\,\footnote{\,The
singular $p$-dependence $\Gamma_2\sim p^{2-\eta}$ arises
near the phase transitions points where $m^2= 0$. For small
$m^2$, when the correlation radius $\xi$ is finite,
$p$-dependence remains singular for $p \gg \xi^{-1}$,
but it is regular for $p \ll \xi^{-1}$.  In the present case,
$m^2$ is finite and in fact tends to infinity in the strong
coupling limit; so the $p$-dependence is surely regular for small
$p$.}
$$
K_2(p)=K_2-\tilde K_2 p^2+\ldots
\eqno(45)
$$
Expressions for the $Z$-factors, renormalized mass and charge
follows from (7)
$$
Z=\left[\frac{\partial}{\partial p^2} \Gamma_2(p)
\right]^{-1}_{p=0} = \frac{K_2^2}{K_0 \tilde K_2} \,,
\eqno(46)
$$
$$
m^2=Z\Gamma_2(0)=\frac{K_2}{\tilde K_2}   \,,
\eqno(47)
$$
$$
g=m^{-\epsilon} Z^2 \Gamma_4=
-\left(\frac{K_2}{\tilde K_2} \right)^{d/2}
\frac{K_4 K_0}{K_2^2}     \,,
\eqno(48)
$$
$$
\frac{1}{Z_2}=\Gamma_{12}\{p_i=0\}=
\frac{d \Gamma_2(0)}{d m_0^2}
=\left(\frac{K_0}{ K_2} \right)'
=\frac{K'_0  K_2-K_0  K'_2}{ K_2^2} \,,
$$
and
$$
\frac{d m^2}{d m_0^2} =\left(\frac{K_2}{\tilde K_2} \right)'
=\frac{K'_2 \tilde K_2-K_2 \tilde K'_2}{\tilde K_2^2}\,,
\eqno(49)
$$
where the prime denotes derivatives over $m_0^2$.
As in Sec.3, parameters  $g_0$ and $\Lambda $ are
considered to be fixed; then $m^2$ is a function of only
$m_0^2$ and derivative $d m_0^2/d m^2$ is determined by
the expression, inverse to (49). Using definition (9) for RG
functions, we have
$$
\beta(g)=\left(\frac{K_2}{\tilde K_2} \right)^{d/2}
\left\{ -d \frac{K_4 K_0}{K_2^2}
+2 \frac{(K'_4 K_0+K_4 K'_0)K_2 -2K_4 K_0 K'_2}{K_2^2}
\frac{\tilde K_2}{K_2\tilde K'_2-K'_2\tilde K_2 }
\right\}
\eqno(50)
$$
$$
{    }
$$
$$
\eta(g)=-\frac{2 K_2\tilde K_2}{K_2\tilde K'_2-K'_2\tilde K_2 }
\left[ 2 \frac{K'_2}{K_2}
 - \frac{K'_0}{K_0} - \frac{\tilde K'_2}{\tilde K_2}
\right]
\eqno(51)
$$
$$
{    }
$$
$$
\eta_2(g)=\frac{2 K_2\tilde K_2}{K_2\tilde K'_2-K'_2\tilde K_2 }
\left\{\frac{  K''_0  K_2 - K_0 K''_2 }
{K'_0 K_2-K_0  K'_2 }
 -2 \frac{ K'_2}{ K_2}
\right\}
\eqno(52)
$$
Eqs.(48), (50), (51), (52) determine $\beta(g)$, $\eta(g)$,
$\eta_2(g)$ in the parametric form: for fixed $g_0$ and $\Lambda$,
the right hand sides of these equations are the functions of only
$m_0^2$, while dependence on the specific choice of $g_0$ and
$\Lambda$ is absent due to general theorems (Sec.2).

If we make  the change of variables $\varphi\to\varphi
(u/8)^{-1/4}$, then all integrals are convergent for any
value of $t\sim g_0^{-1/2}$ (see (24)) and hence they are
regular in the whole comlex $t$ plane except of $t=\infty$.
Any infinities in the right hand sides of Eqs.\,48,\,50--52
can be related only with the zeroes of functional
integrals.\,\footnote{\,It is well-known
from the phase transitions theory,
that singularities of functional integrals can be related only
with the points  $m^2=0$.  In this case, the
correlation radius $\xi$ is infinite and we really need to have
an infinite system and make the singular thermodynamic limit.
If $m^2\ne 0$, then $\xi$ is finite and we can take
the system size ${\cal L}$ much larger than $\xi$ but finite.
If the condition ${\cal L}\gg \xi\gg \Lambda^{-1}$ is
fulfilled, the functional integrals can be approximated by
finite-dimentional ones  and have no singularities for finite
$t$.}
It is clear from Eq.(48) that the limit $g\to\infty$ can be
achieved by two ways: tending to zero either $K_2$, or
$\tilde K_2$. For $\tilde K_2\to 0$, equations  (48) and (50--52)
give
$$
g=-\left(\frac{K_2}{\tilde K_2} \right)^{d/2}
\frac{K_4 K_0}{K_2^2}     \,, \qquad
\beta(g)=- d \left(\frac{K_2}{\tilde K_2} \right)^{d/2}
\frac{K_4 K_0}{K_2^2}
\,,\qquad
\eta(g)\to 2\,,\qquad \eta_2(g)\to 0
\eqno(53)
$$
and the parametric representation is resolved in the form
$$
\beta(g)=d g\,,\qquad \eta(g)=2\,,\qquad \eta_2(g)= 0\,\qquad
(g\to \infty)\,.
\eqno(54)
$$
For $K_2\to 0$, the limit $g\to\infty$ can be achieved
only for $d<4$:
$$
\beta(g)=(d-4) g\,,\qquad \eta(g)=4\,,\qquad \eta_2(g)\to 4
\,\qquad (g\to \infty)\,. \eqno(55)
$$
The results (54), (55)  correspond probably to the different
branches of the function  $\beta(g)$. It is easy to understand
that the physical branch is the first of them. Indeed, it
is commonly accepted in phase transitions theory that
properties of $\varphi^4$ theory change smoothly
as a function of space dimension,
and results for
$d=2,3$ can be obtained by analytic continuation from
$d=4-\epsilon$. All available information indicates on
positiveness of $\beta(g)$  for  $d=4$ (Sec.1), and consequently
its asymptotics at $g\to\infty$ is positive; the same property is
expected for $d<4$ by continuity. The result (54) does obey such
property, while the branch (55) does not exist for $d=4$ at all.
Eq.54 agrees with the approximate results mentioned in Sec.1 and
with the exact result $\beta(g)=2g$ for the
asymptotics of $\beta$-function of the 2D Ising model
\cite{23}, obtained from the duality relation\,\footnote{\,
Definition
of the $\beta$-function in \cite{23} differs by the sign from
the present paper.}. For $d=0$, Eq.54 does not agree with (31)
by the reasons discussed in Sec.3.

\vspace{6mm}
\begin{center}
{\bf 5. Concluding remarks}
\end{center}

According to above considerations,
the standard renormalization procedure defines  theory
for $0\le g \le g_{max}$,  where $g_{max}$ is finite. For
values $g_{max}< g  <\infty$, the theory is defined by analytic
continuation, and large values of
$g$ correspond to complex values of $g_0$.
The latter situation looks inadmissible: the $S$-matrix
can be expressed through the Dyson $T$-exponential of
the bare action, and Hermiticity of the bare Hamiltonian
looks crucial for unitarity of theory.

In fact, a situation is more complicated, as demonstrated by
Bogolyubov's axiomatical construction of the $S$-matrix \cite{2}:
according to it, the general form of the $S$-matrix is given by
the $T$-exponential of $iA$, where $A$ is a sum of (i) the bare
action, and (ii) a sequence of arbitrary "integration constants"
which are determined by quasi-local operators.  In the regularized
theory we can set the "integration constants" to be zero, and the
$S$-matrix is determined by the bare action. However, in the
course of renormalization these constants are taken non-zero, in
order to remove divergences (in fact, Bogolyubov's theorem on
renormalizability is based on this construction).  These non-zero
"integration constants" can be absorbed by the action due to the
change of its parameters. As a result, for the true continual
theory the $S$-matrix  is determined by the renormalized action.
Physically, it is quite reasonable because the bare Hamiltonian
does not exist and the Schr$\ddot o$dinger equation is
ill-defined. From this point of view there is no problem with the
complex bare parameters, since the renormalized Lagrangian is
Hermitian for real $g$.\,\footnote{\,The bare Hamiltonian should
be taken Hermitian in the process of renormalization, since
perturbation theory is different for non-Hermitian operators.
When the relation of renormalized parameters with bare
ones is obtained, it can be analytically continued; possibility
of such continuation is clear from Eqs.47,48.}

Some problems remain for regularized theory, where the bare and
renormalized Lagrangians are equally admissible and a situation
looks controversial: the renormalized Lagrangian is Hermitian and
corresponds to  unitary theory, while the bare Lagrangian is
non-Hermitian and unitarity looks spoiled. The analogous situation
was discussed for the exactly solvable Lee model \cite{xx1}, which
also has the complex bare coupling for the sufficiently large
renormalized coupling. After the paper \cite{xx2} it was generally
accepted and fixed by textbooks \cite{xx3,xx4} that the Lee model
is physically unsatisfactory due to existence of "ghost" states
(i.e. the states with a negative norm). Quite recently \cite{xx5}
it was found that this point of view is incorrect and the Lee
model is completely acceptable physical theory. It is a key idea
of \cite{xx5}  (see also \cite{xx6}) that analytical continuation
of the Hamiltonian parameters to the complex plane should be
assisted by modification of the inner product for the
corresponding Hilbert space. Instead of the usual definition
$$
(f,g)=\int f^*(x) g(x) dx
$$
the inner product is defined as
$$
(f,g)_G= (f, \hat G g) =\int f^*(x) G(x,y) g(y) dx dy
$$
and with the proper choice of the operator $\hat G$ the bare
Hamiltonian is Hermitian in respect to this inner product. As a
result, all states of the Lee model have a positive norm and
evolution is unitary. Modification of the inner product does not
imply any revision of quantum mechanical axioms, if it is applied
to formally defined bare Hamiltonians related with artificial
constructions (like an auxiliary lattice) and not existing in
reality. The interesting question arises, is such modification
admissible in physically relevant situations\,\footnote{\,If the
answer is negative, then large values of renormalized coupling are
inaccessible in the condensed matter applications of $\varphi^4$
theory (where the lattice bare Hamiltonians have a physical
sense).  However, analytical continuation of the $\beta$-function
to arbitrary $g$ is useful from viewpoint of summation of
perturbation series. It should be clear from \cite{4,7} that
knowledge of the strong coupling asymptotics essentially
simplifies the summation procedure and makes it well-defined and
more efficient.  On the other hand, the value $g_{max}$  depends
on the specific lattice and can be large in some cases.}
\cite{xx6}; in fact, arising of non-locality (the kernel $G(x,y)$
instead $\delta(x-y)$) is rather natural consequence of
regularization.


The analogous procedure should exist
in the present case, in order to remove controversy
and give possibility  to define the $\beta$-function and
anomalous dimensions for all positive $g$. This procedure can be
put in a more general context. The arbitrary choice of non-zero
"integration constants" in Bogolyubov's construction allows to
express the  $S$-matrix through different renormalized
Lagrangians: it is a well-known ambiguity of the renormalization
scheme \cite{20b} corresponding to different definitions of
renormalized parameters. Replacement
of one definition of charge
 ($g$) for another ($\tilde g$) corresponds to a
change of variables $g=f(\tilde g)$, not affecting the values
of observable quantities but giving a different parametrization
for them. Let  theory is
satisfactory for some specific definition of $g$; then there
exist a lot of other satisfactory definitions given by "good"
functions $g=f(\tilde g)$.
 Also, there are a lot of definitions
given by "bad" functions (e.g. singular or complex), for which
the theory is looking not satisfactory.
From this point of view,
the paper \cite{xx5} gives a constructive interpretation for a
"bad" change of variables $g=f(g_0)$ (the bare charge in
regularized theory can be consider as a particular definition of
the renormalized charge).


Our results for the $\beta$-function confirm that accepted
definition of the renormalized parameters (Eqs.\,47,\,48)
is satisfactory. Indeed, one can accept the arbitrary
positive value for renormalized charge $g$ at the fixed scale
$\mu$, while its values for other scales are determined by the
Gell-Mann -- Low equation. This definition does not possess any
pathologies like the Landau pole.

Few comments should be given for the dependence of our result
(54)  on the renormalization scheme. From the mathematical side,
the change of variables $g=f(\tilde g)$ is completely arbitrary.
Physically, we should accept some restrictions on it, if we want
that both definitions $g$ and $\tilde g$ were definitions of
"charge", i.e.  give some measure of the vertex $\Gamma_4$. As
the minimal physical restriction we can accept that the function
$ g=f(\tilde g)$ should be regular and give one-to-one
correspondence for physical values of $g$ and $\tilde g$.  If
such restriction is accepted, then the change of variables
$g=f(\tilde g)$  does not allow to transfer between three
qualitatively different situations in the Bogolyubov and Shirkov
classification (see \cite{xx7} for details).  If the "zero
charge" situation is absent for one reasonable definition of
charge, it will be absent for another.

Less rigorously, we can argue that two "physical"
definitions of charge differ in the manner, by which
the vertex $\Gamma_4$
is related with a length scale $L$. If dependence on $L$
has a power-law character, then the order of magnitude
uncertainty in $L$  gives a factor of the
order of unity in definition of charge; so $g \sim \tilde g$.
Then   the linear asymptotics $\beta(g)=\beta_\infty g$
in one case produces the same asymptotics in the other
case, as can be proved by contradiction (since
$g\sim L^{-\beta_\infty}$ for small $L$, then
the difference in $\beta_\infty$ violates the relation
$g \sim \tilde g$).

\vspace{2mm}

\qquad\qquad\qquad\qquad\qquad\qquad----------------------
\vspace{2mm}

In conclusion, the strong coupling asymptotics of the
$\beta$-function in $\varphi^4$ theory is shown to be linear
in the general $d$-dimensional case. In four dimensions,
it means possibility
to construct continuous theory with finite interaction at
large distances.

\vspace{5mm}
The author is indebted for discussions to the participants of L.N.Lipatov's seminar in
Petersburg Nuclear Physics Institute and of B.L.Ioffe's seminar in
Institute of Theoretical and Experimental Physics; he is also grategul to C.M.Bender
for discussion of the Lee model.

\vspace{7mm}


\end{document}